\documentclass[prd,showpacs,twocolumn]{revtex4}
\usepackage{amsmath}
\usepackage{amssymb}
\usepackage{epsfig}
\usepackage{graphicx}
\usepackage{array}
\usepackage{multirow}
\usepackage{color}
\usepackage{txfonts}
\usepackage[colorlinks, citecolor=blue, linkcolor=red, urlcolor=blue]{hyperref}
\allowdisplaybreaks[4]
\begin{document}
\title{Study of $B_s \to \phi (\rho^0, \omega)$ Decays in Standard Model and Family Non-universal $Z^\prime$ Model}
\author{Ying Li} \email{liying@ytu.edu.cn}
\author{Yue Sun}
\author{Zhi-Tian Zou}
\affiliation
{Department of Physics, Yantai University, Yantai 264005,China}
\begin{abstract}
Within QCD factorization approach, we employ the new results of form factors of $B_s \to \phi$ and calculate the branching fractions, CP asymmetries (CPAs) and polarization fractions of the decay modes $B_s \to \phi (\rho^0, \omega)$ in both the standard model (SM) and the family non-universal $Z^\prime$ model. We find that in SM the above observables are relate to two parameters $\rho_H$ and $\phi_H$, which
characterize the end-point divergence in the hard spectator-scattering amplitudes. When setting $\rho_H=0.5$ and $\phi_H\in [-180,180]^\circ$, the theoretical uncertainties are large. By combining the experimental data, our results can be used to constrain these two parameters. Supposing $\rho_H=0$, we study the effects of $Z^\prime$ boson on the concerned observables. With the available branching fraction of $B_s \to \phi \rho^0$, the possible ranges of parameter spaces are obtained. Within the allowed parameter spaces, the branching fraction of $B_s \to \phi \omega$  can be enlarged remarkably. Furthermore, the new introduced weak phase plays important roles in significantly affecting the CPAs and polarization fractions, which are important observables for probing the effects of NP. If these decay modes were measured in the on-going LHC-b and Belle-II experiments in future, the peculiar deviation from SM could provide a signal of the family non-universal $Z^\prime$ model, which can also be used to constrain the mass of $Z^\prime$ boson in turn.
\end{abstract}
\maketitle
\section{Introduction}\label{sec:introduction}
It is well known that the hadronic charmless $B$-meson weak decays play important roles in testing the flavor dynamics of the standard model (SM) and searching for possible effects of new physics (NP) beyond of SM. Particularly, decays dominated by contributions from flavour-changing neutral-currents (FCNC) provide a sensitive probe for NP because their amplitudes are described by loop (or penguin) diagrams where new particles may enter. Moreover, the SM predictions for CP asymmetry (CPA) in several decays are tiny, making them ideal places to look for effects of NP. In past decades, there are already many measurements in rare $B$ decays at the B-factories Belle and BABAR, and the LHC-experiments, such as $B\to \pi \pi$, $B\to \pi K^{*}$, $B\to \phi K^{*}$ and $B_s \to \phi\phi$ decays \cite{ParticleDataGroup:2022pth}.

Although new particles have not been observed directly in the on-going LHC experiments, several experimental results on rare $b$ decays show poor agreement with the corresponding SM predictions. For example, the LHCb collaboration reported deviations in the angular distribution of the $B \to K^*\mu^+ \mu^-$ decay (the ``$P_5^\prime$ anomaly") and in the branching fractions of the $B_s \to \phi \mu^+ \mu^-$, $B \to K^*\mu^+ \mu^-$, $B \to K \mu^+ \mu^-$ and $B\to K^{(*)}\nu\bar \nu$ decays (for recent reviews, see, e.g., \cite{Li:2018lxi, Bifani:2018zmi} and references therein). Theoretically, in order to explain above deviations, on the one hand, we need to calculate all possible corrections in SM including high-order and high-power corrections, and on the other hand we also consider whether these anomalies are due to contributions of NP. It is found that most deviations of semi-leptonic $B$ decays are related to the FCNC $b\to s\mu^+ \mu^-$  transition. If NP is indeed at the origin of the anomalies in $b\to s\mu^+ \mu^-$ transition, it is natural to expect signals in other observables induced by other $b\to s$ transitions, possibly with different realizations though sharing some common features. A natural extension to explore the possible existence of these signals are the charmless hadronic $B$ decays induced by $b\to s q\bar q$ transitions. Based on this strategy, some decays such as $B_s^0 \to K^{(*)}\overline K^{(*)}$ \cite{Alguero:2020xca,Li:2022mtc} and $B_s^0 \to \phi\phi$ \cite{Kapoor:2023txh} have been studied comprehensively. However, unlike semileptonic decays, $B$ meson hadronic decays suffer from larger uncertainties arising from the nonperturbative inputs and corrections from high-order and high-power, so that it is  more difficult to calculate with a high precision. For this reason, we are encouraged to search for observables of some decay modes that have less theoretical uncertainties.

\begin{figure*}[t]
  \includegraphics[width=4.8cm]{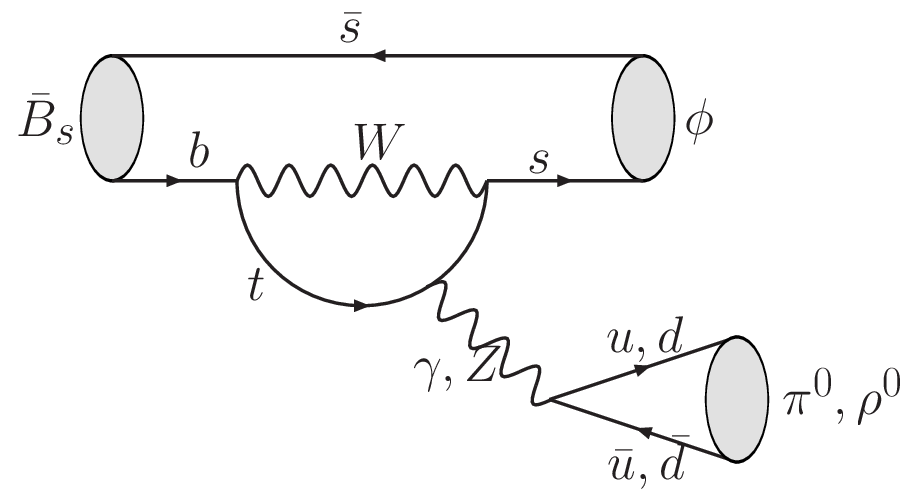}
  \includegraphics[width=4.8cm]{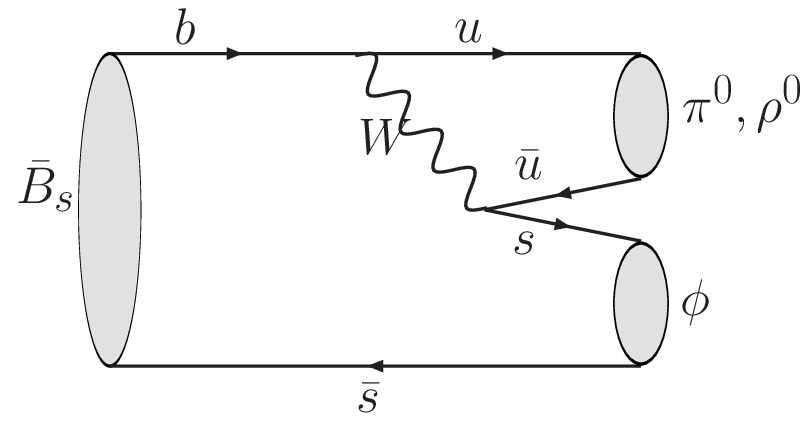}
   \includegraphics[width=4.8cm]{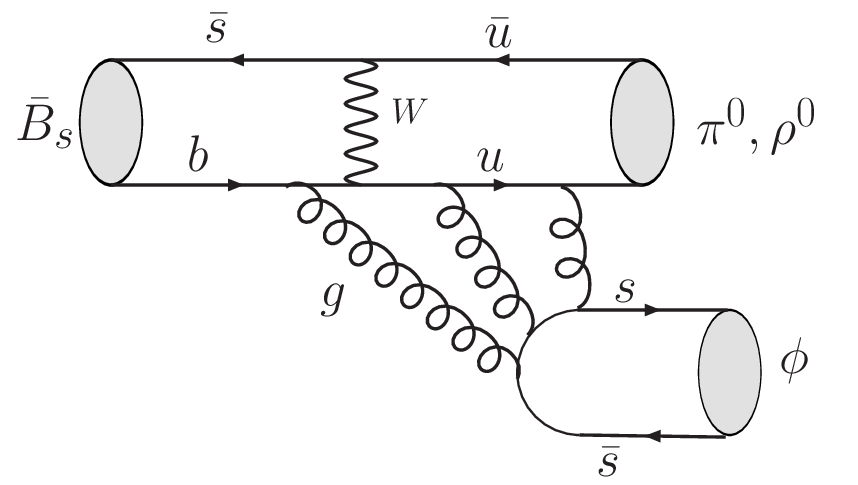}\\
(a)~~~~~~~~~~~~~~~~~~~~~~~~~~~~~~~~~~~~~~~~~~~~~~~~~~~~~~~~~~~~~~
(b)~~~~~~~~~~~~~~~~~~~~~~~~~~~~~~~~~~~~~~~~~~~~~~~~~~~~~~~~~~~~~~
(c)
   \caption{The possible feynman diagrams for the $B_s^0 \to \phi \rho^0$, in which (a) is the electroweak penguin diagram, (b) is the suppressed tree diagram and(c) is the singlet-annihilation diagram. }
  \label{ampsphipi}
\end{figure*}

Motivated by above, we would like to study the process $B_s^0 \to \phi \rho^0$ and $B_s^0 \to \phi \omega$, which are dominated by the FCNC $b\to s q\bar q$ transition. In SM, the possible feynman diagrams for the $B_s^0 \to \phi \rho^0$ are presented in Figure.\ref{ampsphipi}, where Fig.~\ref{ampsphipi}(a) is the electroweak penguin diagram, Fig.~\ref{ampsphipi}(b) is the suppressed tree diagram and Fig.~\ref{ampsphipi}(c) is the singlet-annihilation diagram. In particular, the last diagram called hairpin diagram is viewed as suppressed by Okubo-Zweig-Iizuka (OZI) rules, because the $\rho^0$ meson is produced from at least three gluons, thus this contribution is neglected. Therefore, $B_s^0 \to \phi \rho^0$ decay is independent of weak annihilation contributions and only sensitive to hard spectator scattering corrections, making it an ideal channel for determining the end-point parameters in the hard spectator scattering (HSS) amplitudes. Furthermore, the interference between Fig.~\ref{ampsphipi}(a) and Fig.~\ref{ampsphipi}(b) might lead to possible CPA in this decay.

In order to calculate the two body charmless non-leptonic $B$ decays, several attractive QCD-inspired approaches, such as QCD factorization (QCDF) \cite{Beneke:1999br,Beneke:2000ry}, perturbative QCD (pQCD) \cite{Keum:2000ph,Lu:2000em} and soft-collinear effective theory (SCET) \cite{Bauer:2000yr, Bauer:2001yt}, have been proposed in the last decades. In previous studies, the branching fractions of these decays are shown to be about $10^{-7}$ in SM, both in QCDF \cite{Beneke:2006hg,Chang:2017brr,Cheng:2009mu} and in PQCD \cite{Ali:2007ff}. In the experimental side, the branching fraction of $B_s \to \phi \rho^0 $ decay has been measured by the LHCb collaboration \cite{LHCb:2016vqn},
\begin{eqnarray} \label{expdata}
{\cal B}(B_s \to \phi \rho^0) = (2.7 \pm 0.7_{stat.} \pm 0.2_{ syst.}) \times 10^{-7},
\end{eqnarray}
with a significance of about $4\sigma$. However, the polarization fractions and CPA have not been measured till now. In addition, $B_s \to \phi \rho^0$  has been suggested as a tool to measure $\gamma$ via the mixing-induced CP asymmetry \cite{Fleischer:1994rs}. Since in the era of LHCb and Belle-II these two processes become interesting objects for tests of isospin-violation and potential NP \cite{Belle-II:2018jsg} we will in the following study their phenomenology in full detail, in SM and beyond. In the calculation, we will adopt the QCDF approach, since there is no annihilation contribution in these decays.

Although most experimental data is consistent with the SM predictions, most of us believe that SM is just an effective theory of a more fundamental one yet to be discovered. The presence of a $Z^\prime$ boson associated with an additional $\mathrm{U}(1)^\prime$ gauge symmetry is a well-motivated extension of SM. It should be emphasized that this additional symmetry has not been invented to solve a particular problem of SM, but rather occurs as a byproduct in many models like e.g. grand unified theories, various models of dynamical symmetry breaking and little-Higgs models. An extensive review about the physics of $Z^\prime$ gauge-bosons can be found in \cite{Langacker:2008yv}. The most interesting feature is that the family non-universal $Z^\prime$ couplings could lead to FCNC in the tree level \cite{Buchalla:1995dp,Nardi:1992nq}. The phenomenological effects of $Z^\prime$ in $B$ decays have been studied extensively \cite{Barger:2009qs,Cheung:2006tm,Chang:2009wt,Hua:2010wf,Hua:2010we}. In this work, we will address the effect of the $Z^\prime$ in the rare decay modes $B_s \to \phi \rho^0$ and $B_s \to \phi \omega$. Because these two decays are all penguin dominated process and  mediated by  $b\to s q\bar q$, they are expected to be sensitive to the effect of the $Z^\prime$.

The layout of this paper is as follows. In Sec.\ref{sec:2} we will present the theoretical predictions of $B_s \to \phi \rho^0$ and $B_s \to \phi \omega$ in SM based on QCDF.  Sec.III is devoted to the contributions of $Z^\prime$. Some discussions are also given in this section. We will summarize this work at last.

\section{Predictions in SM} \label{sec:2}
In SM, the effective weak Hamiltonian mediating FCNC transition of the type $b\to s q\bar q~(q=u,d)$ has the form \cite{Buchalla:1995vs}:
\begin{multline} \label{ewhamilton}
 {\cal H}_{eff}={\frac{G_F}{\sqrt 2} }\Big[\sum_{p=u,c}V_{pb}V^*_{ps}\Big(
 C_1O_1^p+C_2O_2^p\Big)\\
 -V_{tb}V^*_{ts}\sum_{i=3}^{10}  C_iO_i\Big],
\end{multline}
where $G_F$ is Fermi coupling constant and $V_{pb}V^*_{ps}$ is the product of the CKM matrix element. $C_i$ are the  Wilson coefficients with the renormalization scale $\mu\sim{m_b}$. The $O_i$ are the local four fermion operators, and $O_{1,2}^p$ are the left-handed current-current operators, $O_{3,...,6}$ and $ O_{7,...,10}$ are the QCD and electroweak penguin operators, respectively, and  they can be expressed as follows:
\begin{eqnarray}
 && O_1^p=(\bar{s}b)_{V-A}(\bar{q}p)_{V-A},  \nonumber \\
 && O_2^p=(\bar{s}_{\alpha}b_{\beta})_{V-A}(\bar{q}_{\beta}p_{\alpha})_{V-A},\nonumber \\
 && O_3=(\bar{s}b)_{V-A}\sum\limits_{q'}(\bar{q}q)_{V-A}, \nonumber \\
 && O_4=(\bar{s}_{\alpha}b_{\beta})_{V-A}\sum\limits_{q}(\bar{q}_{\beta}q_{\alpha})_{V-A},\nonumber \\
 && O_5=(\bar{s}b)_{V-A}{\sum\limits_{q'}}(\bar{q}q)_{V+A}, \nonumber \\
 && O_6=(\bar{s}_{\alpha}b_{\beta})_{V-A}{\sum\limits_{q}}(\bar{q}_{\beta}q_{\alpha})_{V+A},\nonumber \\
 && O_7=(\bar{s}b)_{V-A}\sum\limits_{q}\frac{3}{2}e_{q}(\bar{q}q)_{V+A},\nonumber \\
 && O_8=(\bar{s}_{\alpha}b_{\beta})_{V-A}\sum\limits_{q}\frac{3}{2}e_{q}(\bar{q}_{\beta}q_{\alpha})_{V+A},\nonumber \\
 && O_9=(\bar{s}b)_{V-A}\sum\limits_{q}\frac{3}{2}e_{q}(\bar{q}q)_{V-A},\nonumber \\
 && O_{10}=(\bar{s}_{\alpha}b_{\beta})_{V-A}\sum\limits_{q}\frac{3}{2}e_{q}(\bar{q}_{\beta}q_{\alpha})_{V-A},
\end{eqnarray}
where $\alpha$ and $\beta$ are color indices and $e_{q}$ are charges of the corresponding quark.

In QCDF approach, the hadronic matrix element of the decay ${\overline B}_s \to \phi \rho^0$ can be written as
\begin{multline}
 \langle \phi \rho^0|O_i| \overline B_s^0\rangle = \sum_{j}F_{j}^{B_s^0\to \phi }
 \int_{0}^{1}dx T_{ij}^{I}(x)\Phi_{\rho}(x)  \\
+\int_{0}^{1}d\xi\int_{0}^{1}dx\int_{0}^{1}dy T_{i}^{II}(\xi,x,y)\Phi_{B_s}(\xi)\Phi_{\phi}(x)\Phi_{\rho}(y),
\end{multline}
where $T_{ij}^{I}$ and $T_{i}^{II}$ are the perturbative short-distance interactions and can be calculated perturbatively. $\Phi_{X}(x)~(X=B_s,\pi,\phi)$ are the universal and non-perturbative distribution amplitudes, which can be estimated by the non-perturbative approaches, such as the light cone QCD sum rules, QCD sum rules or lattice QCD.

Because the initial state heavy meson has spin 0, the two vector mesons must have the same helicity due to the conservation of angular momentum. Within the framework of QCDF, the effective Hamiltonian matrix elements are written in the form
\begin{eqnarray}
\left\langle{\phi\rho^0}\left|\mathcal{H}_{eff}\right|\overline{B}_s^0\right\rangle=\frac{G_F}{\sqrt{2}}\sum_{p=u,c}\lambda_p^{(q)}
\left\langle{\phi\rho^0}\left|\mathcal{T}_\mathcal{A}^{h,p}\right|\overline{B}_s^0\right\rangle
\end{eqnarray}
where the superscript $h$ denotes the helicity of the final-state meson. ${\mathcal{T}_\mathcal{A}^{h,p}}$ describes contributions from naive factorization, vertex corrections, penguin contractions and spectator scattering expressed in terms of the flavor operators $a_i^{p,h}$. Specifically,

\begin{eqnarray}
\mathcal{T}_\mathcal{A}^{p,h}
&=&a_1^{p,h}(M_1M_2)\delta_{pu}(\bar{s}b)_{V-A}\otimes(\bar{u}u)_{V-A}\nonumber\\
&+&a_2^{p,h}(M_1M_2)\delta_{pu}(\bar{u}b)_{V-A}\otimes(\bar{s}u)_{V-A}\nonumber\\
&+&a_3^{p,h}(M_1M_2)\sum(\bar{s}b)_{V-A}\otimes(\bar{q}q)_{V-A}\nonumber\\
&+&a_4^{p,h}(M_1M_2)\sum(\bar{q}b)_{V-A}\otimes(\bar{s}q)_{V-A}\nonumber\\
&+&a_5^{p,h}(M_1M_2)\sum(\bar{s}b)_{V-A}\otimes(\bar{q}q)_{V-A}\nonumber\\
&+&a_6^{p,h}(M_1M_2)\sum(-2)(\bar{q}b)_{S-P}\otimes(\bar{s}q)_{S+P}\nonumber\\
&+&a_7^{p,h}(M_1M_2)\sum(\bar{s}b)_{V-A}\otimes{\frac{3}{2}}e_{q}(\bar{q}q)_{V+A}\nonumber\\
&+&a_8^{p,h}(M_1M_2)\sum(-2)(\bar{q}b)_{S-P}\otimes{\frac{3}{2}}e_{q}(\bar{s}q)_{S+P}\nonumber\\
&+&a_9^{p,h}(M_1M_2)\sum(\bar{s}b)_{V-A}\otimes{\frac{3}{2}}e_{q}(\bar{q}q)_{V-A}\nonumber\\
&+&a_{10}^{p,h}(M_1M_2)\sum(\bar{q}b)_{V-A}\otimes{\frac{3}{2}}e_{q}(\bar{s}q)_{V-A}\label{qcdfca}
\end{eqnarray}
where $(\bar{q}_1 q_2)_{V\pm A}\equiv \bar{q}_1\gamma_{\mu}(1\pm{\gamma}_5)q_2$ and $(\bar{q}_1 q_2)_{S\pm P}\equiv \bar{q}_1(1\pm{\gamma}_5)q_2$, and the summation is over $q=u,d$. The symbol $\otimes$ indicates that the matrix elements of the operators in $\mathcal{T}_\mathcal{A}$ are to be evaluated in the factorized form.

The decay constant and form factors is defined by \cite{Cheng:2009mu}:
\begin{eqnarray}
\langle V(p,\varepsilon)|V_\mu|0\rangle& =& f_Vm_V\varepsilon^*_\mu,  \\
\langle V(p,\varepsilon)|V_\mu|B(p_B)\rangle & =& \frac{2}{m_B+m_V}\,\epsilon_{\mu\nu\alpha \beta}\varepsilon^{*\nu}p_B^\alpha p^\beta V(q^2),   \\
\langle V(p,\varepsilon)|A_\mu|B(p_B)\rangle & =&i\Bigg\{(m_B+m_V)\varepsilon^*_\mu A_1(q^2)  \nonumber\\
&-&\frac{\varepsilon^*\cdot p_B} {m_B+m_V}\,(p_B+p)_\mu A_2(q^2)\nonumber \\
&-& 2m_V\frac{\varepsilon^*\cdot p_B}{q^2}\,q_\mu\big[A_3(q^2)-A_0(q^2)\big]\Bigg\},
\end{eqnarray}
where $q=p-p'$, $A_3(0)=A_0(0)$, and
\begin{eqnarray}
A_3(q^2)=\frac{m_B+m_V}{ 2m_V}\,A_1(q^2)-\frac{m_B-m_V}{2m_V}\,A_2(q^2).
\end{eqnarray}
In this work, the form factors of $B_s\to \phi$ we adopted are from \cite{Bharucha:2015bzk} based on the light-cone sum rules.

\begin{widetext}
For a decay $B(p_B) \to V_1(\varepsilon_{1}^*,p_1)V_2(\varepsilon_{2}^*,p_2)$, its factorizable matrix elements are thus given as
\begin{eqnarray}
X_h^{(BV_1,V_2)}&\equiv&\left\langle{V_2}\left|J_{\mu}\right|0\right\rangle\left\langle{V_1}\left|J^{\mu^\prime}\right|B\right\rangle \\ &=&-if_{V_2}m_2\left[(\varepsilon_{1}^*\cdot\varepsilon_{2}^*)(m_B+m_{V_1})A_1^{BV_1}(m_{V_2}^2) -(\varepsilon_{1}^*\cdot{p_B})(\varepsilon_{2}^*\cdot{p_B})\frac{2A_2^{BV_1}(m_{V_2}^2)}{m_B+m_{V_1}}
+i\epsilon_{\mu\nu\alpha\beta}\varepsilon_{2}^{*\mu}\varepsilon_{1}^{*\nu}p_B^{\alpha}p_1^{\beta}\frac{2V^{BV_1}(m_{V_2}^2)}{(m_B+m_{V_1})}\right],
\end{eqnarray}
where $V_1$ takes the spectator quark of the $B$ meson and $V_2$ is the emitted meson. So, the longitudinal $(h=0)$ and transverse $(h=\pm)$  components are obtained as
\begin{eqnarray}
X_0^{(BV_1,V_2)}&=&\frac{if_{V_2}}{2m_{V_1}}\left[(m_{B}^2-m_{V_1}^2-m_{V_2}^2)(m_{B}+m_{V_1})A_1^{BV_1}(q^2)
-\frac{4{m_B^2}p_c^2}{m_{B}+m_{V_1}}A_2^{B{V_1}}(q^2)\right],\label{X0}\\
X_{\pm}^{(BV_1,V_2)}
&=&-if_{V_2}m_{B}m_{V_2}\left[\left(1+\frac{m_{V_1}}{m_{B}}\right)A_1^{B{V_1}}(q^2)\mp\frac{2p_c}{m_{B}+m_{V_1}}V^{B{V_1}}(q^2)\right].\label{Xpm}
\end{eqnarray}

In QCDF, $a_i^{p,h}$ in Eq.(\ref{qcdfca}) are basically the Wilson coefficients in conjunction with short-distance nonfactorizable corrections including vertex corrections and hard spectator interactions, and they have the expressions as
\begin{eqnarray}
a_i^{p,h}(V_1V_2)=\left(C_i+\frac{C_{i\pm1}}{N_c}\right)N_i^h(V_2)+\frac{C_{i\pm1}}{N_c}\frac{C_{F}\alpha_{s}}{4\pi}
\left[V_i^h(V_2)+\frac{4\pi^{2}}{N_c}H_i^h(V_1V_2)\right]+P_i^{h,p}(V_2),
\end{eqnarray}
where $i=1,...10$. The upper (lower) signs apply when $i$ is odd (even), $C_i$ are the Wilson coefficients, $C_F=(N_c^2-1)/(2N_c)$ with $N_c=3$. The functions $V_i^h(V_2)$ stand for vertex corrections, $H_i^h(V_1V_2)$ for hard spectator interactions with a hard gluon exchange between the emitted meson and the spectator quark and $P_i(V_2)$ for penguin contractions. In addition, the expression of the quantities $N_i^h$ reads
\begin{eqnarray}
N_i^h(V_2)=\Bigg\{
\begin{array}{rl}
0, &  i=6,8\\
1, &  {\rm else}.
\end{array}
\end{eqnarray}
\end{widetext}

\begin{table*}[hbt!]
\begin{center}
\caption{Summary of input parameters }
\label{parameter}
\begin{tabular}{c c c c c c c}
 \hline \hline
B meson parameters\\
\multicolumn{1}{c}{$B$}&\multicolumn{1}{l}{$m_B({\rm GeV})$}&\multicolumn{1}{c}{$\tau_{B}(ps)$}&\multicolumn{1}{c}{$f_B({\rm MeV})$}&\multicolumn{1}{r}{$\lambda_{B}({\rm MeV})$}\\
$B_s$  &$5.366$ &$1.472$ &$230\pm20$ &$300\pm100$\\
\hline
Light vector mensons\\
 \multicolumn{1}{c}{$V$}&\multicolumn{1}{c}{$f_V({\rm MeV})$}&\multicolumn{1}{c}{$f_{V}^{\perp}({\rm MeV})$}&\multicolumn{1}{c}{$a_1^V$}
 &\multicolumn{1}{c}{$a_2^V$}&\multicolumn{1}{l}{$a_1^{\perp,V}$}&\multicolumn{1}{c}{$a_2^{\perp,V}$}\\
$\rho$  &$216\pm3$ &$165\pm9$ &$0$ &$0.15\pm0.07$ &$0$ &$0.14\pm0.02$\\
$\phi$  &$233\pm5$ &$191\pm9$ &$0$ &$0.18\pm0.05$ &$0$ &$0.14\pm0.02$\\
$\omega$  &$197\pm3$ &$148\pm9$ &$0$ &$0.15\pm0.07$ &$0$ &$0.14\pm0.02$\\
\hline
Form factors at $q^2=0$\\
\multicolumn{1}{c}{$A_1^{B_s\phi}$}&\multicolumn{1}{c}{$A_2^{B_s\phi}$}&
\multicolumn{1}{c}{$V^{B_s\phi}$}\\
 $0.296\pm0.03$  &$0.25\pm0.03$  &$0.387\pm0.03$\\
 \hline
Quark masses\\
\multicolumn{1}{c}{$m_b(m_b)({\rm GeV})$}&\multicolumn{1}{c}{$m_c(m_b)({\rm GeV})$}&\multicolumn{1}{c}{$m_c^{pole}/m_b^{pole}$}&\multicolumn{1}{c}{$m_s(2.1 {\rm GeV})({\rm GeV})$}\\
$4.2$  &$0.91$ &$0.3$ &$0.095\pm0.020$\\
\hline
Wolfenstein parameters\\
\multicolumn{1}{c}{$A$}&\multicolumn{1}{c}{$\lambda$}&\multicolumn{1}{c}{$\bar {\rho}$}&\multicolumn{1}{c}{$\bar{\eta}$}&\multicolumn{1}{c}{$\gamma$}\\
$0.825$  &$0.2265$ &$0.1598$ &$0.3499$&$67.6$\\
\hline\hline
\end{tabular}
\end{center}
\end{table*}

It is noted that there are end-point divergences when we study power corrections in QCDF. When we calculate the hard spectator interactions $H_i^h(V_1V_2)$ at twist-3 order, soft and collinear divergences arise from the soft spectator quark \cite{Beneke:1999br}. Since the treatment of end-point divergences is model dependent, this subleading power corrections generally can be studied only in a phenomenological way. In this work, we shall follow \cite{Beneke:1999br, Beneke:2006hg} to model the end-point divergence in $H_i^h(V_1V_2)$ as
\begin{eqnarray}
X_H \equiv\int_{0}^{1}\frac{dx}{1- x}=\left(1+\rho_{H}e^{i\phi_H}\right)\ln\left(\frac{m_B}{\Lambda_h}\right),
\end{eqnarray}
where $\Lambda_h= 500~\rm MeV$ is a typical scale, and $\rho_{H},\phi_{H}$ are the unknown real parameters.

The  amplitudes of $\overline{B}_{s}^0\to{\phi\rho^0}$ and $\overline{B}_{s}^0\to{\phi\omega}$ decays are written as
\begin{multline}
\mathcal{A}_h(\overline{B}_{s}^0 \to\phi\rho^0)=\frac{G_F}{2}\sum_{p=u,c}\lambda_p^{(s)}\Bigg[\delta_{pu}a_2^{p,h}\\ +\frac{3}{2}(a_7^{p,h}+a_9^{p,h})\Bigg] X_h^{(\overline{B}_{s}^0\phi, \rho)};
\end{multline}
\begin{multline}
\mathcal{A}_h(\overline{B}_{s}^0 \to\phi\omega) =\frac{G_F}{2}\sum_{p=u,c}\lambda_p^{(s)}\Bigg[\delta_{pu}a_2^{p,h}\\
+2(a_3^{p,h}+a_5^{p,h})+\frac{1}{2}(a_7^{p,h}+a_9^{p,h})\Bigg] X_h^{(\overline{B}_{s}^0\phi, \omega)}.
\end{multline}
We note that the transverse amplitudes $\mathcal{A}_\pm$ are suppressed by a factor $m_2/m_B$ relative to $\mathcal{A}_0$, as shown in Eqs.(\ref{X0}) and (\ref{Xpm}). In addition, the axial-vector and vector contributions to $\mathcal{A}_+$ are cancelled by each other in the heavy-quark limit, due to an exact form factor relation \cite{Beneke:2000wa}. Thus, in quark model \cite{Korner:1979ci} or naive factorization \cite{Kramer:1991xw}, the hierarchy of helicity amplitudes
\begin{eqnarray}
\mathcal{A}_0:\mathcal{A}_-:\mathcal{A}_+=1:\frac{\Lambda_{\rm QCD}}{m_b}:\left(\frac{\Lambda_{\rm QCD}}{m_b}\right)^2
\end{eqnarray}
is therefore  expected. This hierarchy can also be explained by the chirality flip \cite{Kagan:2004uw}. The transverse amplitudes defined in the transversity basis are related to the helicity ones via
\begin{eqnarray}
\mathcal{A_\Vert}=\frac{\mathcal{A}_{+}+\mathcal{A}_{-}}{\sqrt{2}},
\mathcal{A_\perp}=\frac{\mathcal{A}_{+}-\mathcal{A}_{-}}{\sqrt{2}}.
\end{eqnarray}

\begin{figure*}[!htb]
  \includegraphics[width=5.7cm]{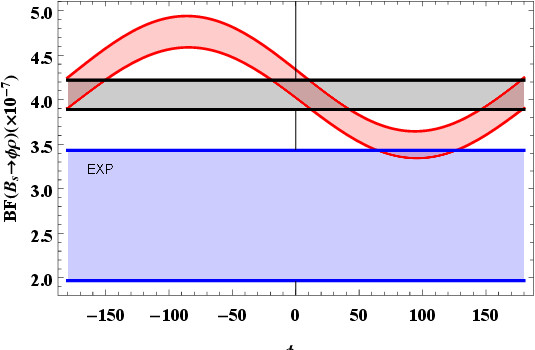}
  \includegraphics[width=5.7cm]{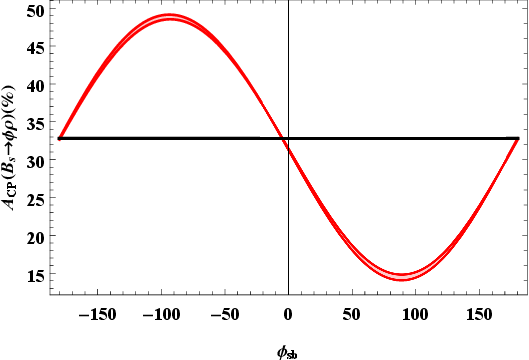}
  \includegraphics[width=5.7cm]{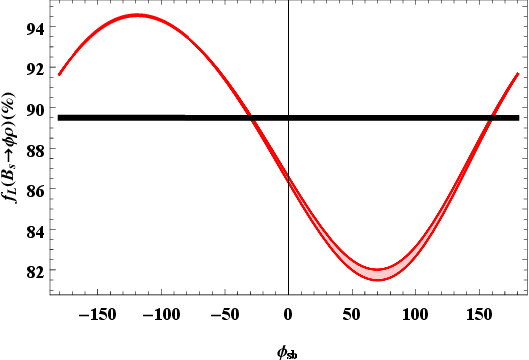}
   \caption{The possible regions of CP averaged branching fraction, CPA and longitudinal polarization fraction of decay $B_s\to \phi\rho^0$ as a function of the phase $\theta_H$, the red regions represent results from the $\rho=0.5$, and the horizontal (grey) regions are results of $\rho=0$.}
  \label{Bsphirho}
\end{figure*}
\begin{figure*}[!htb]
  \includegraphics[width=5.7cm]{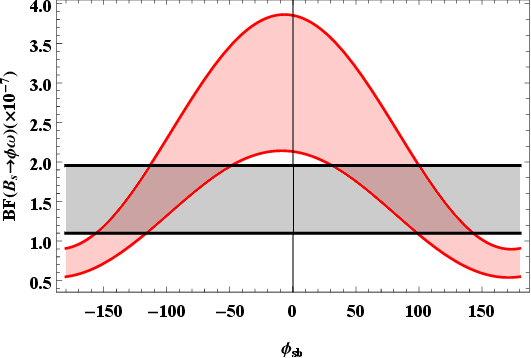}
  \includegraphics[width=5.7cm]{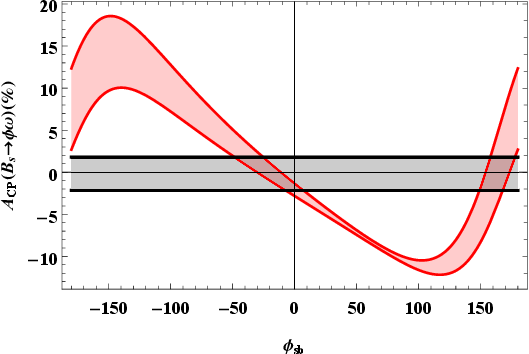}
  \includegraphics[width=5.7cm]{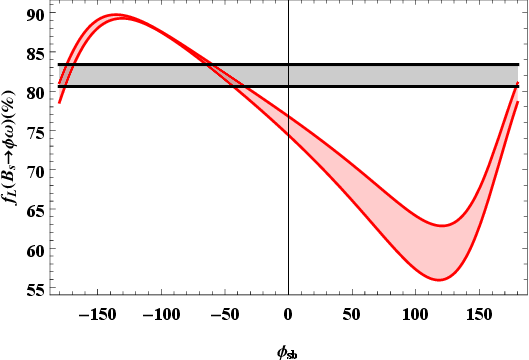}
   \caption{The possible regions of CP averaged branching fraction, CPA and longitudinal polarization fraction of decay $B_s\to \phi\omega$ as a function of the phase $\theta_H$, the red regions represent results from the $\rho=0.5$, and the horizontal (grey) regions are results of $\rho=0$.}
  \label{Bsphiomega}
\end{figure*}

With the amplitudes, we then obtain the branching fraction of $\overline B_s^0\to V_1V_2$ as
\begin{eqnarray}
\mathcal{B}(\overline B_s^0\to V_1V_2)=\frac{\tau_{B_s}\left|p_c\right|}{8\pi{m_{B_s}}^2}
\left(\left|\mathcal{A}_{L}\right|^{2}+\left|\mathcal{A}_{\Vert}\right|^{2}+\left|\mathcal{A}_{\perp}\right|^{2}\right),
\end{eqnarray}
where $\tau_{B_s}$ is the lifetime of the $B_s$ meson and $\left|p_c\right|$ is the absolute value of two final-state hadrons’ momentum in the $B_s$ rest frame. Then, three polarization fractions are then defined as
\begin{eqnarray}
f_{\alpha}\equiv\frac{\left|\mathcal{A}_{\alpha}\right|^2}{\left|\mathcal{A}_{0}\right|^{2}
+\left|\mathcal{A}_{\Vert}\right|^{2}+\left|\mathcal{A}_{\perp}\right|^{2}}
\end{eqnarray}
with $\alpha=L,\Vert,\perp$. In addition, we can also define the direct CPAs as:
\begin{eqnarray}
A_{CP}=\frac
{|\mathcal{A}(\overline{B}_s\to \overline{V}_1\overline {V}_2)|^2-|\mathcal{A}(B_s\to V_1V_2)|^2}
{|\mathcal{A}(\overline{B}_s\to \overline{V}_1\overline {V}_2)|^2+|\mathcal{A}(B_s\to V_1V_2)|^2};\\
A_{CP}^\alpha=\frac
{f_\alpha(\overline{B}_s\to \overline{V}_1\overline {V}_2)-f_\alpha(B_s\to V_1V_2)}
{f(_\alpha\overline{B}_s\to \overline{V}_1\overline {V}_2)+f_\alpha(B_s\to V_1V_2)}.
\end{eqnarray}

Using the input parameters listed in Table.~\ref{parameter}, we now calculate the branching fractions, polarization fractions and direct CPAs. As aforementioned, two of the most important parameters are $\rho_H$ and $\phi_H$. We firstly follow \cite{Beneke:2006hg} and adopt the default values $\rho_H=0$ and $\phi_H=0$. Three observables of decay $B_s\to \phi\rho^0$ and $B_s\to \phi\omega$ in SM are presented as
\begin{eqnarray}
&&\left\{
  \begin{array}{l}
    {\cal B}(B_s\to \phi\rho^0)=(4.1\pm 0.5)\times 10^{-7},   \\
    {\cal B}(B_s\to \phi\omega)=(1.5^{+0.6}_{-0.5})\times 10^{-7};  \\
     \end{array}
\right.\\
&&\left\{
  \begin{array}{l}
    f_L(B_s\to \phi\rho^0)=(89.5^{+0.9}_{-1.0})\%,   \\
    f_L(B_s\to \phi\omega)=(82.1^{+2.4}_{-2.8})\%; \\
     \end{array}
\right.\\
&&\left\{
  \begin{array}{l}
     A_{CP}(B_s\to \phi\rho^0)=(32.8^{+0.1}_{-0.6})\%,   \\
     A_{CP}(B_s\to \phi\omega)=(-0.6^{+3.8}_{-2.1})\%;  \\
     \end{array}
\right.
\end{eqnarray}
where the errors are from the uncertainties of the nonperturbative parameters. It can be seen that the longitudinal polarization fractions and the CPAs are not sensitive to the nonperturbative inputs, such as decay constants, form factors, and moments in the distribution amplitudes. In addition, we find that for $B_s \to \phi\rho^0$ the major uncertainties are from the form factors, while the decay constants dominate the uncertainties in $B_s \to \phi\omega$. The branching fractions are in agreement with the previous predictions of \cite{Beneke:2006hg}, and are larger than those of \cite{Cheng:2009mu}, because the form factors of $B_s\to \phi$ we used are larger than theirs. Comparing to the experimental data \cite{LHCb:2016vqn} shown in Eq.~(\ref{expdata}), our results is a bit larger than the data. The other observables have not been measured till now.

In ref.\cite{Chang:2017brr}, the authors had fitted $\rho_H=0.5$  within the all $B_{u,d,s}\to VV$ data. Within these ranges, we suppose $\phi_H\in[-180^\circ, 180^\circ]$ and obtain the results as
\begin{eqnarray}
&&\left\{
  \begin{array}{l}
    {\cal B}(B_s\to \phi\rho^0)=(4.2^{+0.7+0.6}_{-0.6-0.5})\times 10^{-7},   \\
    {\cal B}(B_s\to \phi\omega)=(2.9^{+1.1+0.0}_{-0.9-2.2})\times 10^{-7},  \\
     \end{array}
\right.\\
&&\left\{
  \begin{array}{l}
    f_L(B_s\to \phi\rho^0)=(86.5^{+1.0+8.0}_{-2.2-5.3})\%,   \\
    f_L(B_s\to \phi\omega)=(75.7^{+2.3+13.4}_{-2.7-16.1})\%,  \\
     \end{array}
\right.\\
&&\left\{
  \begin{array}{l}
     A_{CP}(B_s\to \phi\rho^0)=(31.4^{+0.8+15.6}_{-2.8-14.7})\%,   \\
     A_{CP}(B_s\to \phi\omega)=(-2.3^{+1.6+14.3}_{-0.8-8.9})\%,  \\
     \end{array}
\right.
\end{eqnarray}
where the first errors are from the uncertainties of the nonperturbative parameters, and the second ones come from the phase $\phi_H$. From the results, it is obvious that the strong phase $\phi_H$ takes larger uncertainties, especially for decay $B_s \to \phi \omega$. Moreover, when the $\phi_H$ is in the range $[35,160]^\circ$, the branching fraction of $B_s \to \phi\rho^0$ agrees with the data \cite{LHCb:2016vqn}.  In order to show the effects of the parameters $(\rho_H, \phi_H)$, we set $\rho_H=0$ and $\rho_H=0.5$ and plot the variations of the branching fractions, CPAs and longitudinal polarization fractions of $B_s\to \phi\rho^0$ and $B_s\to \phi\omega$ decays with the phase $\phi_H$ in Fig.\ref{Bsphirho} and Fig.\ref{Bsphiomega}, respectively. In fact, if $\rho_H$ becomes larger, the uncertainties taken by $\phi_H$ will be even larger. In all figures, only uncertainties of the form factors (decay constants) in $B_s \to \phi\rho^0$ ($B_s \to \phi\omega$) are considered.  In future, if these above observables could be measured with high precision, our theoretical results will be useful for determining the range of $(\rho_H, \phi_H)$ in turn.

\section{Effects of New Physics}
\begin{table*}[t]
 \begin{center}
 \caption{The Wilson coefficients $C_i$ within the SM and with the contribution
 from $Z^{\prime}$ boson included in NDR scheme at the scale $\mu=m_{b}$ and
 $\mu_h=\sqrt{\Lambda_h m_b}$ \cite{Chang:2009wt}.}
 \label{Wilson}
 \vspace{0.1cm}
 \doublerulesep 1.0pt \tabcolsep 0.04in
 \begin{tabular}{c|cc|cc}\hline\hline
 Wilson                   &\multicolumn{2}{c|}{$\mu=m_{b}$}       &\multicolumn{2}{c}{$\mu_h=\sqrt{\Lambda_h m_b}$} \\
 \cline{2-3}\cline{4-5}
 coefficients             &$C_i^{SM}$ &$\Delta C_i^{Z^{\prime}}$  &$C_i^{SM}$ &$\Delta C_i^{Z^{\prime}}$\\ \hline\hline
 $C_1$                    &$1.075$    &$-0.006\xi^L$      &$1.166$    &$-0.008\xi^L$\\
 $C_2$                    &$-0.170$   &$-0.009\xi^L$      &$-0.336$   &$-0.014\xi^L$ \\
 $C_3$                    &$0.013$    &$0.05\xi^L-0.01\xi^R$  &$0.025$    &$0.11\xi^L-0.02\xi^R$\\
 $C_4$                    &$-0.033$   &$-0.13\xi^L+0.01\xi^R$ &$-0.057$   &$-0.24\xi^L+0.02\xi^R$\\
 $C_5$                    &$0.008$    &$0.03\xi^L+0.01\xi^R$  &$0.011$    &$0.03\xi^L+0.02\xi^R$\\
 $C_6$                    &$-0.038$   &$-0.15\xi^L+0.01\xi^R$ &$-0.076$   &$-0.32\xi^L+0.04\xi^R$\\
 $C_7/{\alpha}_{em}$      &$-0.015$   &$4.18\xi^L-473\xi^R$   &$-0.034$   &$5.7\xi^L-459\xi^R$\\
 $C_8/{\alpha}_{em}$      &$0.045$    &$1.18\xi^L-166\xi^R $  &$0.089$    &$3.2\xi^L-355\xi^R$\\
 $C_9/{\alpha}_{em}$      &$-1.119$   &$-561\xi^L+4.52\xi^R$  &$-1.228$   &$-611\xi^L+6.7\xi^R$\\
 $C_{10}/{\alpha}_{em}$   &$0.190$    &$118\xi^L-0.5\xi^R$    &$0.356$    &$207\xi^L-1.4\xi^R$\\
  \hline \hline
 \end{tabular}
 \end{center}
 \end{table*}

Though when $\rho_H=0.5$, the SM prediction of branching fraction of $B_s \to \phi\rho^0$ can explain the data, we have chances to search for the effects of NP, because  $\rho_H$ and $\phi_H$ have not completely confirmed yet. In this section, we will study the effects of an extra $Z^\prime$ gauge boson in these decays $B_{s} \to \phi (\rho^0,\omega)$. Ignore the mixing between $Z^0$ and $Z^\prime$, we write the couplings of the $Z^\prime$ to fermions as \cite{Langacker:2008yv}
\begin{eqnarray}
J_{Z^{\prime}}^{\mu}=g^{\prime}\sum_i \bar\psi_i
\gamma^{\mu}[\epsilon_i^{\psi_L}P_L+\epsilon_i^{\psi_R}P_R]\psi_i,
\label{eq:JZprime}
\end{eqnarray}
where $i$ is the family index and $\psi$ labels the fermions and $P_{L,R}=(1\mp\gamma_5)/2$. In some string and GUT models \cite{Chaudhuri:1994cd,Cleaver:1998gc,Cvetic:2001tj,Cvetic:2002qa,Kuo:1984gz,Barger:1987hh}, the $Z^{\prime}$ couplings are not required to be family universal. When rotating to the physical basis, FCNCs generally appear at tree level in both left handed and right handed sectors, explicitly, as
\begin{eqnarray}
B^{L}=V_{\psi_L}\epsilon^{\psi_L}V_{\psi_L}^{\dagger},\;\;\;\;\;
B^{R}=V_{\psi_R}\epsilon^{\psi_R}V_{\psi_R}^{\dagger}.
\end{eqnarray}
For simplicity, we suppose that the right-handed couplings are flavor-diagonal and $B_{sb}^R=0$. As a result, the $Z^{\prime}$ part of the effective Hamiltonian for $b\to s\bar{q}q\,(q=u,d)$ transition has the form as:
\begin{multline}\label{heffz1}
 {\cal H}_{eff}^{\rm
 Z^{\prime}}=\frac{2G_F}{\sqrt{2}}\big(\frac{g^{\prime}M_Z}
 {g_1M_{Z^{\prime}}}\big)^2
 \,B_{sb}^L(\bar{s}b)_{V-A}\\
 \times \sum_{q}\left[B_{qq}^L (\bar{q}q)_{V-A}
 +B_{qq}^R(\bar{q}q)_{V+A}\right],
\end{multline}
where $g_1=e/(\sin{\theta_W}\cos{\theta_W})$ and $M_{Z^{\prime}}$ is the $Z^\prime$ mass. Compared with the effective weak Hamiltonian of SM shown in Eq.(\ref{ewhamilton}),  the above Hamiltonian Eq.~(\ref{heffz1}) can be rearranged as
\begin{multline}
 {\cal H}_{eff}^{\rm
 Z^{\prime}}=
 -\frac{G_F}{\sqrt{2}}V_{tb}V_{ts}^{\ast}\\
 \times \sum_{q}
 (\Delta C_3 O_3^q +\Delta C_5 O_5^q+\Delta C_7 O_7^q+\Delta C_9
  O_9^q)\,,
\end{multline}
where $O_i^q(i=3,5,7,9)$ are the effective operators of SM. $\Delta C_i$ are the modifications to the corresponding SM Wilson coefficients caused by $Z^{\prime}$ boson, which are expressed as
\begin{eqnarray}
 \Delta C_{3}&=&-\frac{2}{3V_{tb}V_{ts}^{\ast}}\,\big(\frac{g^{\prime}M_Z}
 {g_1M_{Z^{\prime}}}\big)^2\,B_{sb}^L\,(B_{uu}^{L}+2B_{dd}^{L})\,,\nonumber\\
 \Delta C_{5}&=&-\frac{2}{3V_{tb}V_{ts}^{\ast}}\,\big(\frac{g^{\prime}M_Z}
 {g_1M_{Z^{\prime}}}\big)^2\,B_{sb}^L\,(B_{uu}^{R}+2B_{dd}^{R})\,,\nonumber\\
 \Delta C_{7}&=&-\frac{4}{3V_{tb}V_{ts}^{\ast}}\,\big(\frac{g^{\prime}M_Z}
 {g_1M_{Z^{\prime}}}\big)^2\,B_{sb}^L\,(B_{uu}^{R}-B_{dd}^{R})\,,\nonumber\\
 \Delta C_{9}&=&-\frac{4}{3V_{tb}V_{ts}^{\ast}}\,\big(\frac{g^{\prime}M_Z}
 {g_1M_{Z^{\prime}}}\big)^2\,B_{sb}^L\,(B_{uu}^{L}-B_{dd}^{L})\,,
 \label{NPWilson}
\end{eqnarray}
in terms of the model parameters at the $M_W$ scale. It is found that the $Z^\prime$ contributes not only to the EW penguins operators but also to the QCD penguins ones. In particular, we suppose that new physics is manifest in the EW penguins by setting $B_{uu}^{L,R}=-2 B_{dd}^{L,R}$ as done in \cite{Buras:2003dj}. Due to the hermiticity of the effective Hamiltonian, the diagonal elements of the effective coupling matrices $B_{qq}^{L,R}$ are real. However, for the off-diagonal one $B_{sb}^{L}$, it could be a complex with a new weak phase $\phi_{sb}^L$. As a result, the resulting $Z^\prime$ contributions to the Wilson coefficients are then written as:
\begin{eqnarray}
& &\Delta C_{3,5}\simeq 0, \nonumber\\
& &\Delta
C_{9,7}=4\frac{|V_{tb}V_{ts}^{\ast}|}{V_{tb}V_{ts}^{\ast}}\xi_{sb}^{L,R}e^{-i\phi_{sb}^L},
\end{eqnarray}
with
\begin{eqnarray}
\xi_{sb}^{L,R}=\left(\frac{g^{\prime}M_Z}
 {g_1M_{Z^{\prime}}}\right)^2\left|\frac{B_{sb}^LB_{dd}^{L,R}}{V_{tb}V_{ts}^{\ast}}
 \right|.
\end{eqnarray}

The next step is to constrain the ranges of the new defined parameters $\xi_{sb}^{L,R}$. In generally, we suppose that both the $U(1)_Y$ and $U^\prime(1)$ gauge groups origin from the same grand unified theories, and $g^\prime /g_1 \sim 1 $ is expected.  The direct search for $Z^\prime$ boson is one of important physics programs of current and future high-energy colliders. However, the direct signal of the new ${Z^{\prime}}$ boson have not been observed in the current experiments such as CMS and ATLAS, implying that the mass of $Z^\prime$ would be larger than the TeV scale. In this work, we set $ M_{Z^{\prime}}\geq 3 \rm TeV$ conservatively. Because the family non-universal $Z^\prime$ leads to $\Delta B=2$ and $\Delta S=2$ FCNC, so that $B_s^0-\overline{B}_s^0$ mixing happens at the tree level. Then, the mass difference $\Delta m_{B_s}$ presents the most strong constraint to the models with $Z^\prime$ boson, and $\left|B_{sb}^{L}\right| \sim\left|V_{tb} V_{ts}^{*}\right|$ is theoretically required \cite{Chang:2009tx, Li:2012xc, Barger:2004qc}. Meanwhile, in order to explain CPAs of $B \to K \pi$ and branching fractions of $B \to K \phi$ and $B \to K^* \phi$, the diagonal elements should satisfy $\left|B_{ss}^{L,R} \right|\sim\left|B_{dd}^{L,R}\right|\sim \mathcal{O}(1)$ \cite{Cheung:2006tm, Barger:2009qs}. In addition this parameter region is of interest for collider detection. Based on the above results obtained,  we shall use
\begin{eqnarray}
|\xi|=|\xi_{db}^{L,R}|=|\xi_{sb}^{L,R}|\in(1 \sim 2)\times 10^{-3}.
\end{eqnarray}
We also note that the other SM Wilson coefficients also receive contributions from the $Z^\prime$ boson through renormalization group (RG) evolution. Given that there is no significant RG running effect between $M_{Z^\prime}$ and $M_W$ scale, the RG evolution of the modified Wilson coefficients is exactly the same as the ones in the SM \cite{Buchalla:1995vs}.  The Wilson coefficients at $m_b$ and $\sqrt{\Lambda_h m_b}$ scale have been presented in Table.~\ref{Wilson}.

\begin{figure*}[!htb]
  \includegraphics[width=5.7cm]{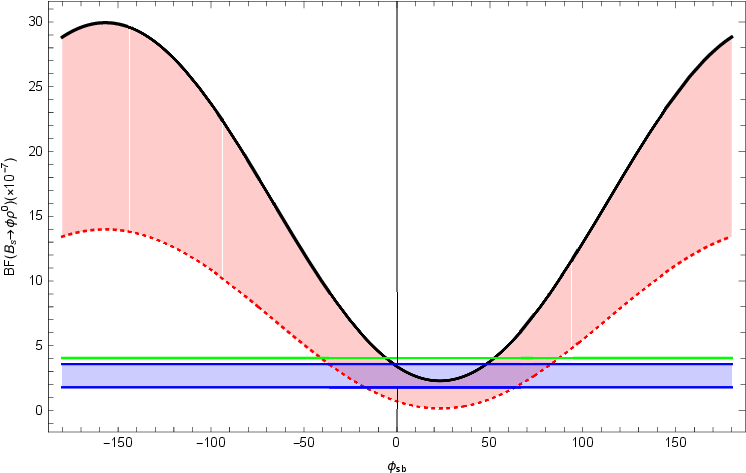}
  \includegraphics[width=5.7cm]{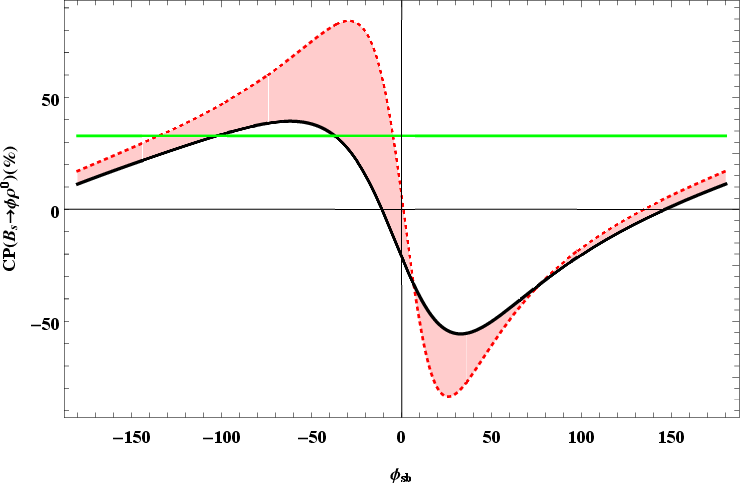}
  \includegraphics[width=5.7cm]{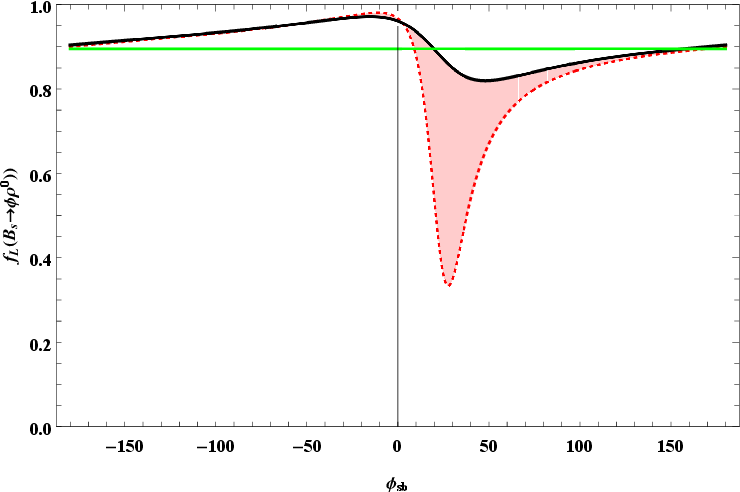}
   \caption{The CP averaged branching fractions, CPAs and longitudinal polarization fractions of decay $B_s\to \phi\rho^0$ as a function of the new weak phase $\phi_{sb}$, the dotted (red),solid (black) lines represent results from the $\xi= 0.001, 0.002$, and the solid lines the horizontal (green)lines are the center values of SM.}
  \label{x1}
\end{figure*}
\begin{figure*}[!htb]
  \includegraphics[width=5.7cm]{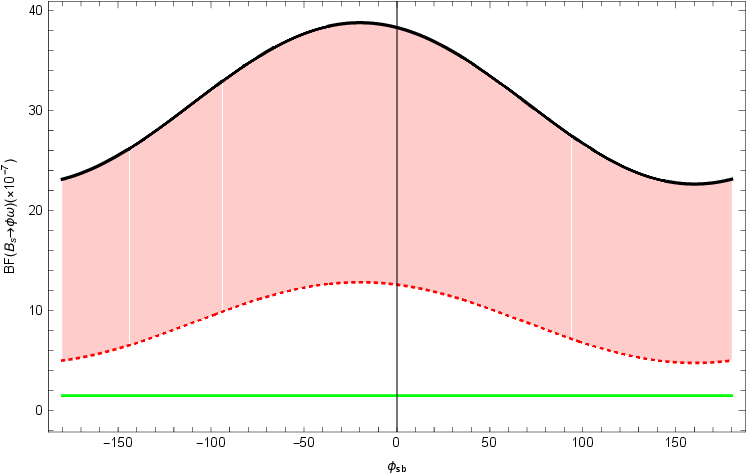}
  \includegraphics[width=5.7cm]{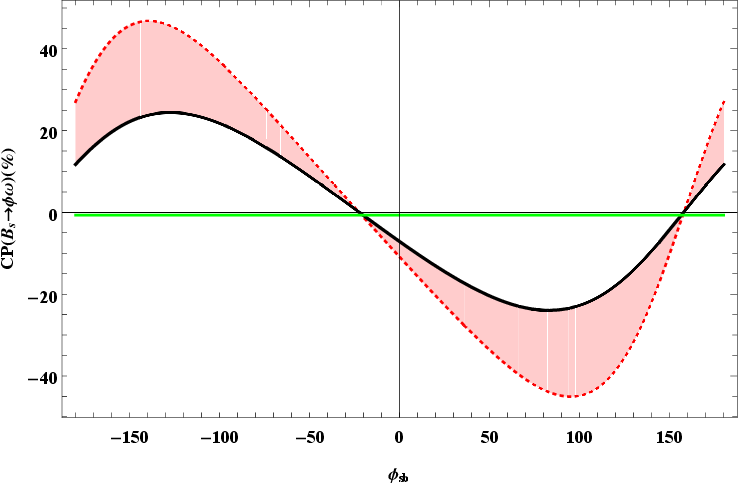}
  \includegraphics[width=5.7cm]{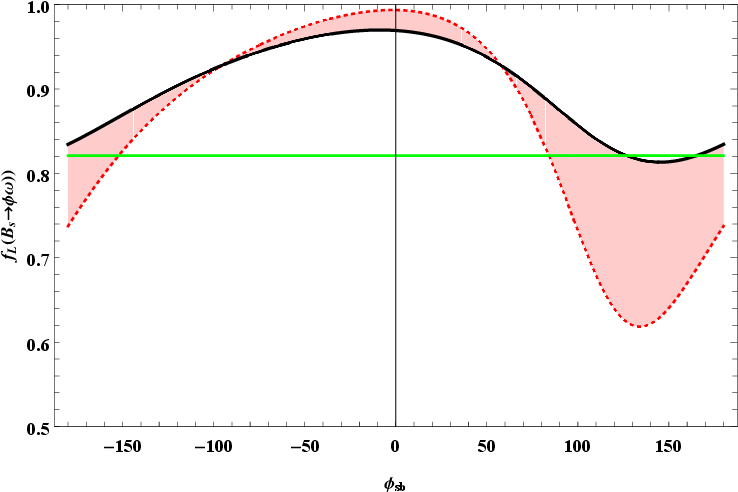}
   \caption{The CP averaged branching fractions, CPAs and longitudinal polarization fractions of decay $B_s\to \phi\omega$ as a function of the new weak phase $\phi_{sb}$, the dotted (red), solid (black) lines represent results from the $\xi= 0.001, 0.002$, and the solid lines the horizontal (green)lines are the center values of SM.}
  \label{x2}
\end{figure*}

To illustrate the effect of the $Z^\prime$ boson, by setting $\rho_H=0$ and $\xi=(1, 2)\times 10^{-3}$, one can get the variations of the CP averaged branching fractions, CPAs and longitudinal polarization fractions of decays $B_s\to \phi\rho^0$ and $B_s\to \phi\omega$ as a function of the new weak phase $\phi_{sb}$, as shown in Fig.~\ref{x1} and Fig.~\ref{x2}, where the horizontal lines are the center values predicted in SM.  From left panel of Fig.~\ref{x1}, we find that large $\xi>0.003$, namely a lighter $Z^\prime$ boson, is ruled out. When $\xi=0.001$, $\phi_{sb} \in(-40, 80)^\circ$ is favored. With this range, the branching fraction of $B_s\to \phi\omega$ would be $(0.8\sim 1.2) \times 10^{-6}$, which is larger than that of SM by one order of magnitude, as shown in left panel of Fig.~\ref{x2}. As aforementioned, the newly introduced weak phase $\phi_{sb}$ in the off-diagonal element of $B_{sb}^{L}$ plays a major role in changing CPA. The CPAs are much sensitive to $\xi$ and $\phi_{sb}$. For $B_s\to \phi\rho^0$, if $\xi=0.002$, the range of CPA is $-56\%\sim 38\%$. If $\xi=0.001$, its CPA could reach $\pm 85\%$ for some special $\phi_{sb}$. The CPA of $B_s\to \phi\omega$ shares the same characteristics with $B_s\to \phi\rho$. These remarkable changes shown in the center panels of Fig.~\ref{x1} and Fig.~\ref{x2}  will be important signals in testing the model.

For the decay $B\to \phi\rho^0$, the longitudinal polarization fraction $f_L$ is not sensitive to nonperturbative parameters if we adopt $\rho_H=0$ and $\phi_H=0$, as shown in the right panel of Fig.~\ref{Bsphirho}. When including the contribution of $Z^\prime$, it could decrease to 0.33, when $\xi=0.001$ and $\phi_{sb}=27^\circ$. If $\phi_{sb}=-10^\circ$, it could be enhanced to $0.97$. For the decay $B\to \phi\omega$, if $\phi_{sb} \in(-40, 80)^\circ$, the $f_L$ is in the range $(0.84, 1)$, which is larger than the prediction of SM. In addition, we also note that when $\xi=0.002$, the changes is not as remarkable as the case of $\xi=0.001$, which can be seen in the right panels of of Fig.~\ref{x1} and Fig.~\ref{x2}. The reason is that,  when $\xi=0.001$, the contribution of $Z^\prime$ is comparable with that of SM, however when $\xi=0.002$, the contribution of $Z^\prime$ becomes larger than that of SM and dominate the amplitude. If we only consider effect of $Z^\prime$, $f_L=0.91$ for both decays. In future, these observables could be used to probe the effect of new physics. If the $Z^\prime$ were detected in the colliders directly, these decays would also be useful to constrain the couplings.

\section{Summary}
In this work, we calculated the branching fractions, CP asymmetries and polarization fractions of the decay mode $B_s \to \phi \rho^0$  and $B_s \to \phi \omega$ within the QCD factorization approach in both the SM and the family non-universal $Z^\prime$ model. This approach is suitable because these decay modes have no contributions from annihilation diagrams. In SM, both decays are sensitive to two parameters $\rho_H$ and $\phi_H$, which are from end-point singularities in the hard spectator scattering.

Using the latest results of form factors of $B_s \to \phi$, we obtained ${\cal B}(B_s \to \phi \rho^0)=(4.1 \pm 0.5)\times 10^{-7}$ with $\rho_H=0$, and is $(4.2^{+0.7+0.6}_{-0.6-0.5})\times 10^{-7}$ with  $\rho_H=0.5$ and  $\phi_H \in [-180,180]^\circ$. These results are a bit larger than the current experimental data. In addition,  the longitudinal polarization fraction  $f_L$ is $89.4\%$ and $86.3\%$ for different values of $\rho_H$. Due to the interference between tree and penguin contributions, the CPA $A_{CP}$ is about $30\%$. The future measurements of $f_L$ and $A_{CP}$ are helpful to determine the parameters ($\rho_H, \phi_H$). The decay $B_s \to \phi \omega$ was also studied.

Because when we adopt $\rho_H=0.5$, the effects of $Z^\prime$ will be buried in the large uncertainties taken from $\phi_H$, we only study its effects in $B_s \to \phi \rho^0$  and $B_s \to \phi \omega$ by setting $\rho_H=0$. In comparison with experimental data, $\xi>0.003$ (light $Z^\prime$) is ruled out. By setting $\xi=0.001$, the range  $\phi_{sb}\in (-40, 80)^\circ$ is favoured. The branching fraction of $B_s \to \phi \omega$ may be enlarged  by one ordr of magnitude by $Z^\prime$ boson within the allowed  parameter space. The $f_L$ of  $B_s \to \phi \rho^0$ decreases to 0.33 with $\phi_{sb}=27^\circ$, and reaches to 0.97 with $\phi_{sb}=-10^\circ$. Furthermore, as the direct CPA is concerned, it can reach $-85\%$ with suitable parameter space. The contribution of $Z^\prime$ to $B_s \to \phi \omega$ was also calculated. All above observables could be measured in the on-going LHC-b experiment and Belle-II, and the future measurements with high precision  will provide a plate to test the non-universal $Z^\prime$ model, and can be used to constrain the mass of the $Z^\prime$ boson in turn.
\section*{Acknowledgment}
This work was supported in part by the National Science Foundation of China under the Grant Nos.~12375089, and the Natural Science Foundation of Shandong province under the Grant No. ZR2022ZD26.

\end{document}